\documentclass[journal]{IEEEtran}
\usepackage{graphicx}

\ifCLASSINFOpdf
\else
\fi
\usepackage[cmex10]{amsmath}
\usepackage{url}

\begin{document}
%
\title{A Balloon-borne Measurement of High Latitude Atmospheric Neutrons Using a LiCAF
Neutron Detector}
%
%
%

\author{Merlin Kole, Yasushi Fukazawa, Kentaro Fukuda, Sumito Ishizu, Miranda Jackson, Tune Kamae, Noriaki Kawaguchi, Takafumi Kawano, M\'ozsi Kiss, Elena Moretti, Maria Fernanda Mu\~noz Salinas, Mark Pearce, Stefan Rydstr\"{o}m, Hiromitsu Takahashi, Takayuki Yanagida

\thanks{Merlin Kole is with KTH - The Royal Institute of Technology, Department of Physics, 10691 Stockholm, Sweden and The Oskar Klein Centre for Cosmoparticle Physics, AlbaNova, 10691 Stockholm, Sweden, e-mail: merlin@particle.kth.se.}
\thanks{M. Jackson, M. Kiss, E. Moretti, M. Pearce and S. Rydstr\"{o}m are with
KTH Royal Institute of Technology, Department of Physics, and the Oskar
Klein Centre for Cosmoparticle Physics, AlbaNova University Centre, 10691
Stockholm, Sweden}
\thanks{Y. Fukazawa, T. Kawano and H. Takahashi are with the Department of Physical Science, Hiroshima University, Hiroshima 739-8526, Japan}
\thanks{K. Fukuda, S. Ishizu and N. Kawaguchi are with the Tokuyama Corporation, Shunan, Yamaguchi, Japan}
\thanks{T. Kamae is with the University of Tokyo, Dept. of Physics, 113-0033 Tokyo, Japan}
\thanks{T. Yanagida is with the Kyushu Institute of Technology, Kitakyushu, Fukuoka, Japan}}
\maketitle

\begin{abstract}
PoGOLino is a scintillator-based neutron detector. Its main purpose is to provide data on the neutron flux in the upper stratosphere at high latitudes at thermal and non-thermal energies for the PoGOLite instrument. PoGOLite is a balloon borne hard X-ray polarimeter for which the main source of background stems from high energy neutrons. No measurements of the neutron environment for the planned flight latitude and altitude exist. Furthermore this neutron environment changes with altitude, latitude and solar activity, three variables that will vary throughout the PoGOLite flight. PoGOLino was developed to study the neutron environment and the influences from these three variables upon it. PoGOLino consists of two Europium doped Lithium Calcium Aluminium Fluoride (Eu:LiCAF) scintillators, each of which is sandwiched between 2 Bismuth Germanium Oxide (BGO) scintillating crystals, which serve to veto signals produced by gamma-rays and charged particles. This allows the neutron flux to be measured even in high radiation environments. Measurements of neutrons in two separate energy bands are achieved by placing one LiCAF detector inside a moderating polyethylene shield while the second detector remains unshielded. The PoGOLino instrument was launched on March 20th 2013 from the Esrange Space Center in Northern Sweden to an altitude of 30.9 km. A description of the detector design and read-out system is presented. A detailed set of simulations of the atmospheric neutron environment performed using both PLANETOCOSMICS and Geant4 will also be described. The comparison of the neutron flux measured during flight to predictions based on these simulations will be presented and the consequences for the PoGOLite background will be discussed.

\end{abstract}


%
\IEEEpeerreviewmaketitle

\vspace{1cm}

\section{Introduction}
%
%
%
%
\IEEEPARstart{H}{igh} energy neutrons are a source of background for balloon-borne and Earth orbiting X-ray detectors making use of materials with a low atomic number, Z. In the case of low Z scintillator materials, signal and background discrimination requires the use of a scintillator materials with dedicated discrimination abilities and very fast read out electronics. Passive shielding from neutrons adds a significant mass to the payload. A remaining irreducible high energy neutron induced background is therefore often unavoidable. For PoGOLite \cite{PoGO}, a balloon-borne hard X-ray polarimeter, such an irreducible background is also present. PoGOLite aims to measure the polarization of X-rays from point sources using a plastic scintillator array. A design which uses plastic scintillators both for detecting the Compton scattering and the photo-absorption location was chosen in order to maximise the effective area for the energy range of $20-100\,\mathrm{keV}$. The drawback of this material choice is a high sensitivity to neutron 
scattering events. The plastic scintillator material used in PoGOLite, EJ-204, in combination with the sampling rate of the electronics of $37.5\,\mathrm{MHz}$ does not allow for a complete photon-neutron differentiation based on pulse shape discrimination. Both passive and active background reduction systems are therefore required.

\begin{figure}[!t]
\centering
\includegraphics[width=2.5in]{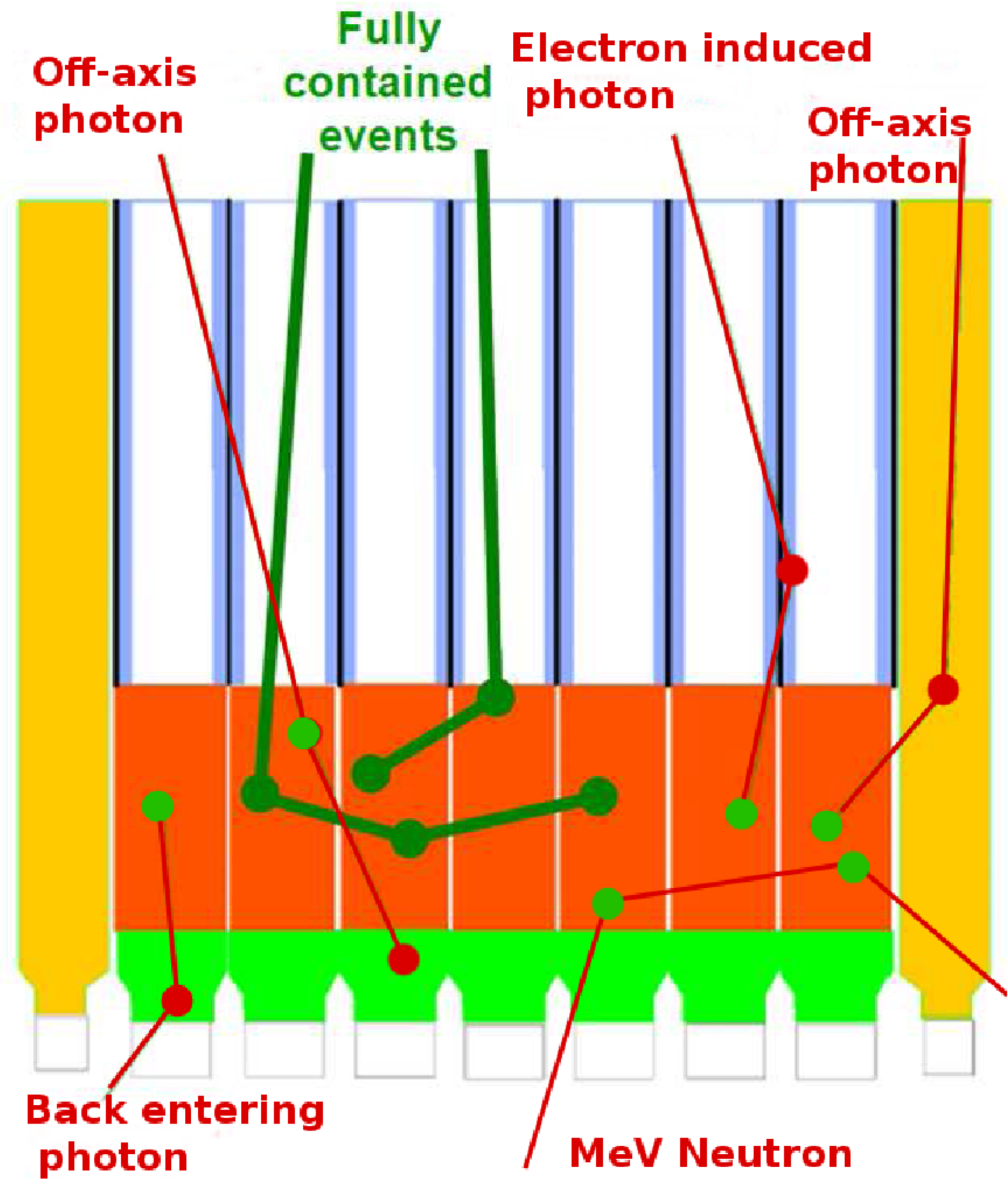}
\caption{A schematic overview of the PoGOLite polarimeter (not to scale). The plastic scintillator array used for signal detection is shown in orange. The active background rejection systems are presented in yellow for the SAS, green for the bottom BGO and blue for the active collimation system. The passive collimators, consisting of lead and tin are shown as black lines between the active collimators. The various types of signal and background events are also presented with a green dot indicating a valid signal like hit and red dots for identified background events.}
\label{fig_bg}
\end{figure}

In PoGOLite several systems are in place to reduce the instrument background to a minimum. The plastic scintillator array, consisting of 61 hexagonal cross-section scintillators with a short decay time, is surrounded by a Side Anti-Coincidence Shield (SAS) consisting of thirty, 60 cm long BGO crystals. On the bottom the plastic scintillator array is shielded by $4\,\mathrm{cm}$ long BGO crystals which are connected to the plastic scintillators and readout using the same photomultiplier tube. In order to restrict the field of view of the instrument, thereby making it suitable for the observation of point sources, each plastic scintillator is connected to a $60\,\mathrm{cm}$ long hollow plastic scintillator. The long tube-like scintillators have a longer decay time than those dedicated for signal detection, and similar to the bottom BGO crystals, pulse shape discrimination is applied to distinguish signal events from background events. Apart from these active background rejection systems each hollow plastic scintillator is wrapped in a thin 
layer of lead and tin which serve as passive collimators. All these systems, of which a schematic representation is shown in figure \ref{fig_bg}, reduce the photon and charged particle background to a minimum. 

Neutrons can enter the plastic scintillator array without leaving a detectable signal in any of the aforementioned background rejection systems. In the plastic scintillator array neutrons can scatter in between plastic scintillators and subsequently leave the instrument. A schematic example of this is shown in figure \ref{fig_bg}. If the energy of the incoming neutron is in the $1-100\,\mathrm{MeV}$ range the scintillator pulse heights resemble those originating from hard X-rays Compton scattering and absorption events in the plastic scintillator. To reduce the incident flux of neutrons in this energy range a $\sim300\,\mathrm{kg}$ heavy polyethylene shield with a maximum thickness of $15\,\mathrm{cm}$ is placed around the polarimeter. This shield serves to slow down high energy neutrons through multiple scatterings towards sub-MeV energies. Neutrons in this energy range will deposit insufficient energy to be mistaken for signal events. Despite the use of the heavy polyethylene shield the predicted background rate is of the same order of magnitude as the expected signal rate from the Crab - the primary observation target for PoGOLite. Increasing 
the shield thickness would result in a higher mass of the payload and thereby reduce the float altitude of the instrument, resulting in a lower detected signal rate. The current thickness of the shield was chosen to optimize the signal to background ratio. 

The neutron energy spectrum and overall flux depend on the altitude, magnetic latitude and solar activity. These three variables will vary through-out the flight of PoGOLite which is circumpolar with a launch from the Esrange Space Centre in Northern Sweden. For this reason the neutron induced background rate will vary between observations. In order to understand the measured signal it is therefore important to understand the atmospheric neutron environment in detail. To achieve this a detailed set of simulations was performed with the aim of providing direction-dependent neutron energy spectra for different altitudes, latitudes and solar activities. Secondly a separate instrument, dedicated to neutron detection, named PoGOLino was designed and launched from the Esrange Space Centre in Northern Sweden. This instrument was designed as a light weight stand alone detector capable of measuring the neutron flux with high statistics in a rapidly changing environment. For PoGOLino this was achieved using LiCAF scintillators which despite their small size have 
a high neutron capturing cross section. 

This paper will provide a brief overview of the atmospheric neutron environment after which the PoGOLino instrument will be described in detail. This is followed by a presentation of the flight results and a comparison with simulations.


\section{Atmospheric Neutrons}

\begin{figure}[!t]
\centering
\includegraphics[width=2.5in]{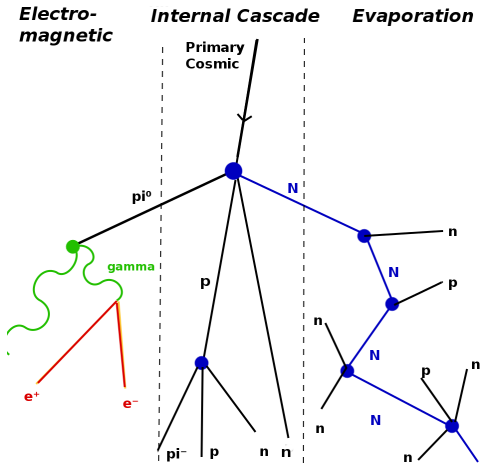}
\caption{A schematic representation of neutron production in a cosmic ray induced air shower. The left side of the figure shows the electromagnetic part of the shower in which no neutrons are produced. The middle shows neutrons produced during an internal cascade and on the right neutron production by nucleon evaporation is visualised.}
\label{air_shower}
\end{figure}

Neutrons are produced in the Earth's atmosphere in air showers induced by charged cosmic rays, the majority of which are protons. The highest energy neutrons are produced by weak interactions between the incoming proton and an atmospheric nucleus. Neutrons produced through this mechanism can have kinetic energies exceeding $1\,\mathrm{GeV}$. Their momentum is directed, like that of the primary cosmic ray, towards the Earth's surface. This production process is however rare due to its small cross section. Most neutrons are produced in head on collisions of cosmic rays with atmospheric nucleons. Initially in such an interaction an internal cascade is formed inside the nucleus \cite{Bert}. Different hadronic particles with energies exceeding the lowest available energy level in the nucleus are produced in this process. This mechanism is responsible for the production of neutrons with energies between $\sim 10\,\mathrm{MeV}$ and $\sim 500\,\mathrm{MeV}$ with momenta directed mostly towards the Earth's surface. After the internal cascade the remaining 
nucleus will be in an excited state from which it will recover by expelling protons, neutrons and photons until reaching the ground state. The neutrons produced through this process, often referred to as evaporation, have energies of $\sim 1\,\mathrm{MeV}$ and have a more isotropic directional dependence. A schematic representation of neutron production through the last two processes in an air shower is shown in figure \ref{air_shower}.

After production neutrons will scatter off atmospheric nuclei, losing energy and changing their direction with each such interaction. As a result the typical atmospheric neutron energy spectrum is a power law ranging from sub-eV energies, occupied by fully thermalised neutrons to GeV energies. Since the particles in the low energy part of the spectrum have undergone more scatterings the momentum direction is more isotropic at lower energies. For altitudes close the the shower maximum, $\sim 15-20\,\mathrm{km}$, the sub-keV part of the spectrum is almost fully isotropic. When moving to higher altitudes, due to the difference in atmospheric density, the majority of the low energy neutrons will move upwards in the atmosphere. 

The dependence of the neutron environment on the magnetic latitude is a result of the modulation of the incoming charged cosmic rays by the Earth's magnetic field. This field shields the equator from protons with energies below $\sim 10\,\mathrm{GeV}$ whereas for a latitude of $60^\circ$ only protons with an energy below $1\,\mathrm{GeV}$ are stopped by the magnetic field. As a result the overall number of primary charged cosmic rays entering the atmosphere increases with magnetic latitude. 

Solar activity influences the neutron environment in the atmosphere due to modulation of the primary cosmic rays. The number of low energy particles ejected by the Sun increases with solar activity. This ejected material interacts with the primary cosmic rays. Therefore a higher solar activity results in a reduction of the number of incoming primary particles in the Earth's atmosphere. The cross section for the interaction of the ejected material with cosmic rays decreases with increasing energy. As a result low energy cosmic rays are more strongly modulated by the Sun. For protons with energies exceeding $\sim20\,\mathrm{GeV}$ the influence of this modulation is not visible. Due to the energy dependence on the modulation the effect of the solar activity is more pronounced at higher latitudes.

The PoGOLite instrument was launched on the 12th of July from Esrange Northern Sweden which has a magnetic latitude of $65^\circ$ degrees. It landed on the 26th of July around $200\,\mathrm{km}$ north of Norilsk in Russia, which has a magnetic latitude of $60^\circ$ degrees. The float altitude varied between $36\,\mathrm{km}$ and $40\,\mathrm{km}$. A more detailed description of the PoGOLite flight of 2013 can be found in \cite{Parky}. Due to the variation of the altitude, latitude and solar activity during the flight a neutron induced background rate that varies from observation to observation is expected. The PoGOLino instrument was launched in order to gain an insight into the neutron environment at similar latitudes during a solar activity similar to that during the PoGOLite flight.


\section{Instrument Design}

\begin{figure}[!t]
  \begin{center}
    \includegraphics[width=0.55\textwidth]{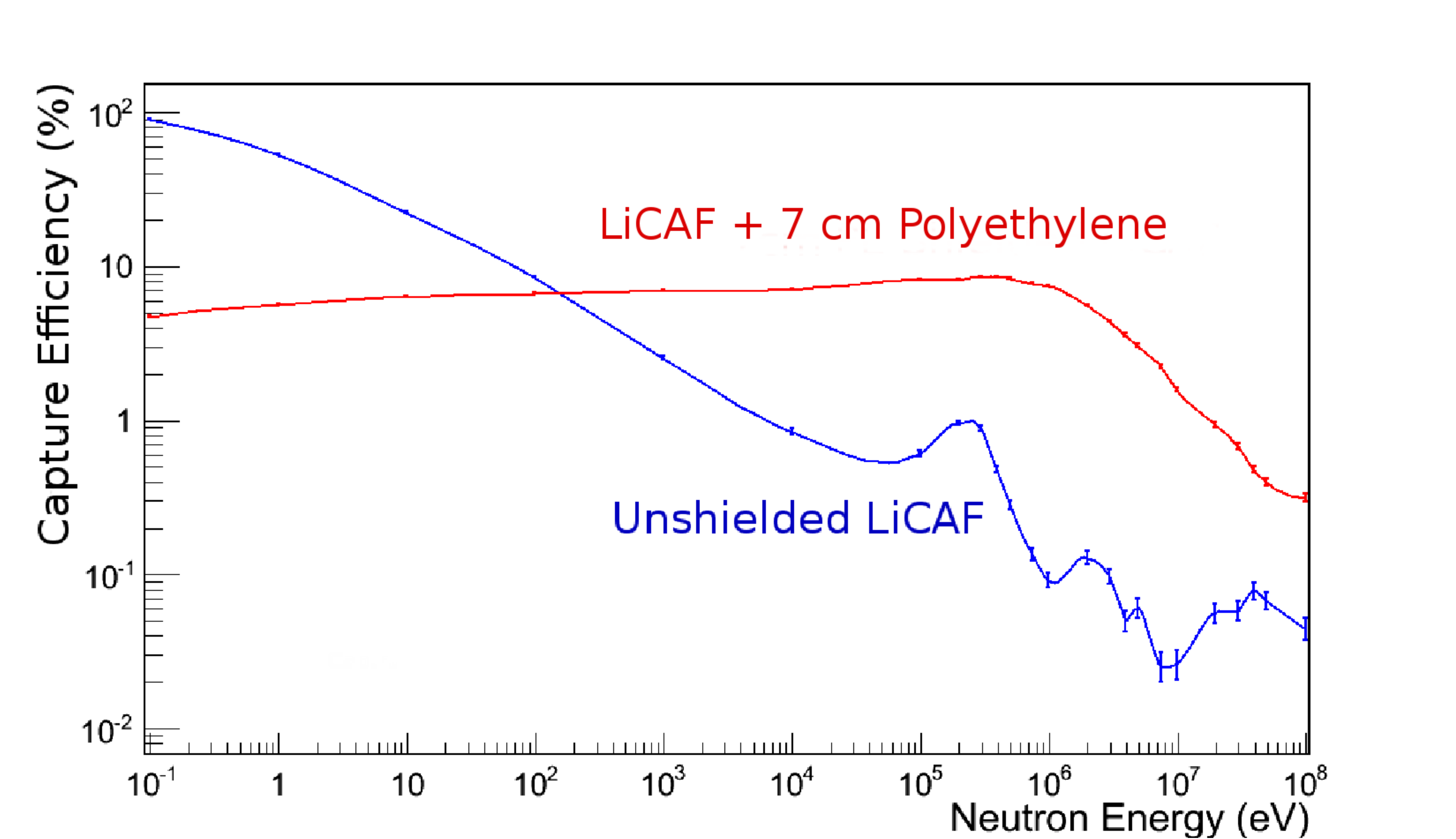}
  \end{center}
  \caption{The neutron capture efficiency as simulated by Geant4 for a bare 5 mm thick sheet of LiCAF (blue) and the same sheet shielded by 70 mm of polyethylene (red).}
  \label{energyresponse}
\end{figure}
\begin{figure}[!b]
  \begin{center}
    \includegraphics[width=0.58\textwidth]{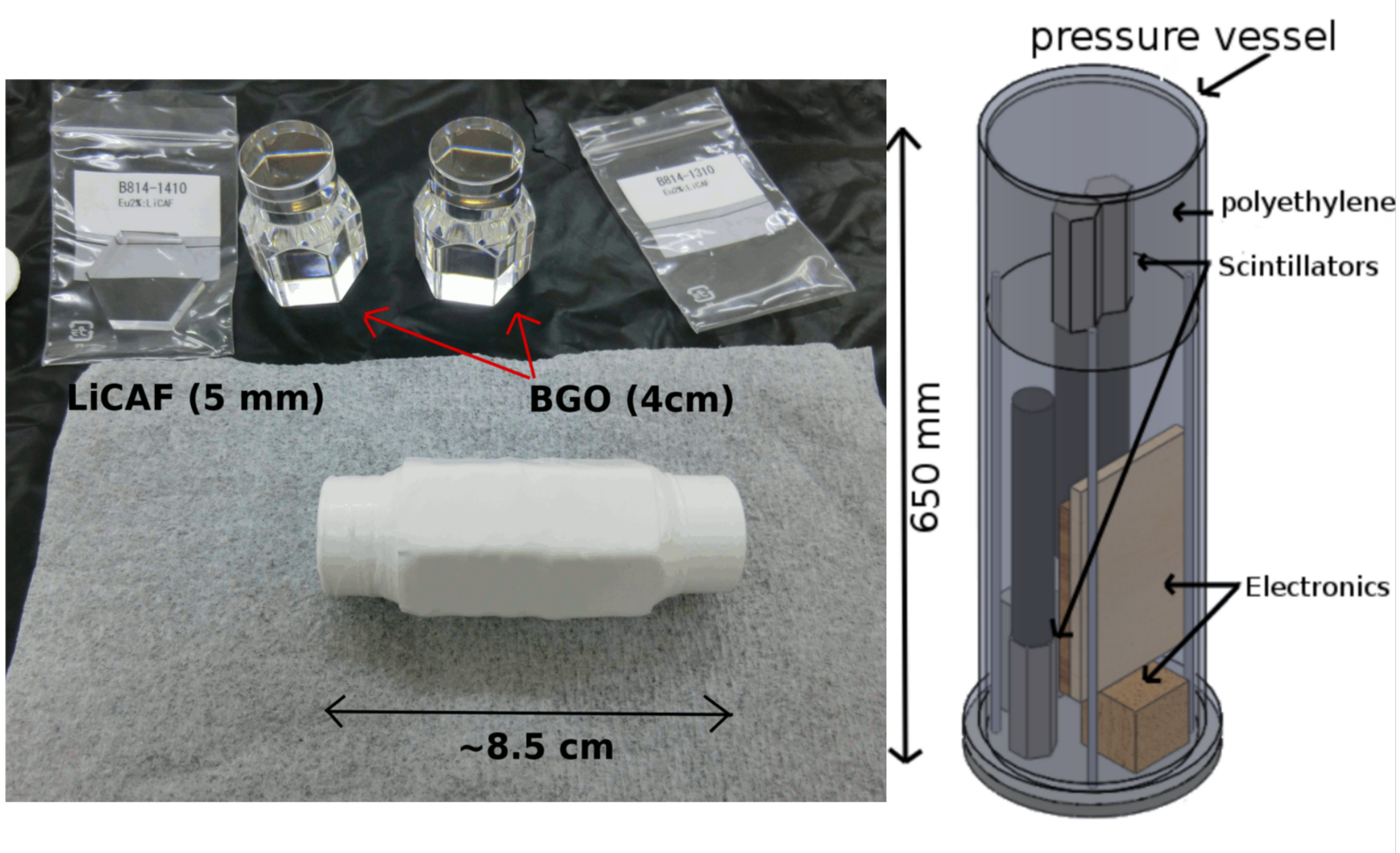}
  \end{center}
  \caption{The photo a neutron scintillator in front of the three individual crystals on the left and a schematic view of the instrument layout on the right.}
  \label{instrument}
\end{figure}

PoGOLino comprises two phoswich type `neutron scintillators', each comprising one hexagonal $5\,\mathrm{mm}$ thick Europium doped Lithium Calcium Aluminium Fluoride (Eu:LiCAF) scintillator sandwiched between two Bismuth Germanium Oxide (BGO) crystals. The different crystals can be seen in figure \ref{instrument}. The lithium in the LiCAF material contains $95\%$ $^6\mathrm{Li}$. $^6\mathrm{Li}$ has a high capture cross section, $\sim 1000\,\mathrm{barn}$, for thermal neutrons. Neutron capture by $^6\mathrm{Li}$ will result in the decay reaction: 

\[^6\mathrm{Li} + \mathrm{n} \rightarrow\,^4\mathrm{He}\,(2.73\,\mathrm{MeV}) +\,^3\mathrm{T}\,(2.05\,\mathrm{MeV})\]

\begin{figure}[!b]
  \begin{center}
    \includegraphics[width=0.40\textwidth]{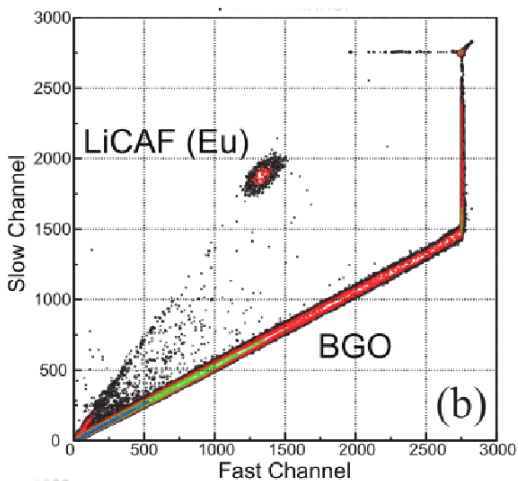}
  \end{center}
  \caption{The "fast" output (relative pulse height at the 4th clock cycle after the trigger) versus the “slow” output (relative pulse height at the 15th clock cycle after the trigger) diagrams of a Eu:LiCAF and BGO phoswich detector irradiated with 252-Cf and 137-Cs. Taken from \cite{HT}.}
  \label{fast_slow}
\end{figure}

Both the high energy charged decay products will deposit their energy in the scintillating material, causing a distinct line in the measured energy spectrum. The two BGO crystals, which have a large cross section for photoelectric absorption, serve to reject high energy gamma-rays. Furthermore they are also used to veto charged particles. The scintillators are read out using a single PMT of the type Hamamatsu R7899EGKNP. Signals originating from the LiCAF can be distinguished from those from the BGO using pulse shape discrimination based on the decay times; for signals from europium doped LiCAF, at room temperature this is $1.6\,\mathrm{\mu s}$ \cite{HT}, for BGO this is $300\,\mathrm{ns}$ \cite{PoGO}. Figure \ref{fast_slow}, taken from \cite{HT}, shows the pulse height measured 4 clock cycles (fast output) after a trigger is issued versus the pulse height measured after 15 clock cycles (slow output) for a Eu:LiCAF BGO scintillator irradiated with $^{137}\mathrm{Cs}$ and $^{252}\mathrm{Cf}$. The distinct neutron capturing peak can clearly be distinguished from the line resulting from BGO events. 

\begin{figure}[!h]
  \begin{center}
    \includegraphics[width=0.50\textwidth]{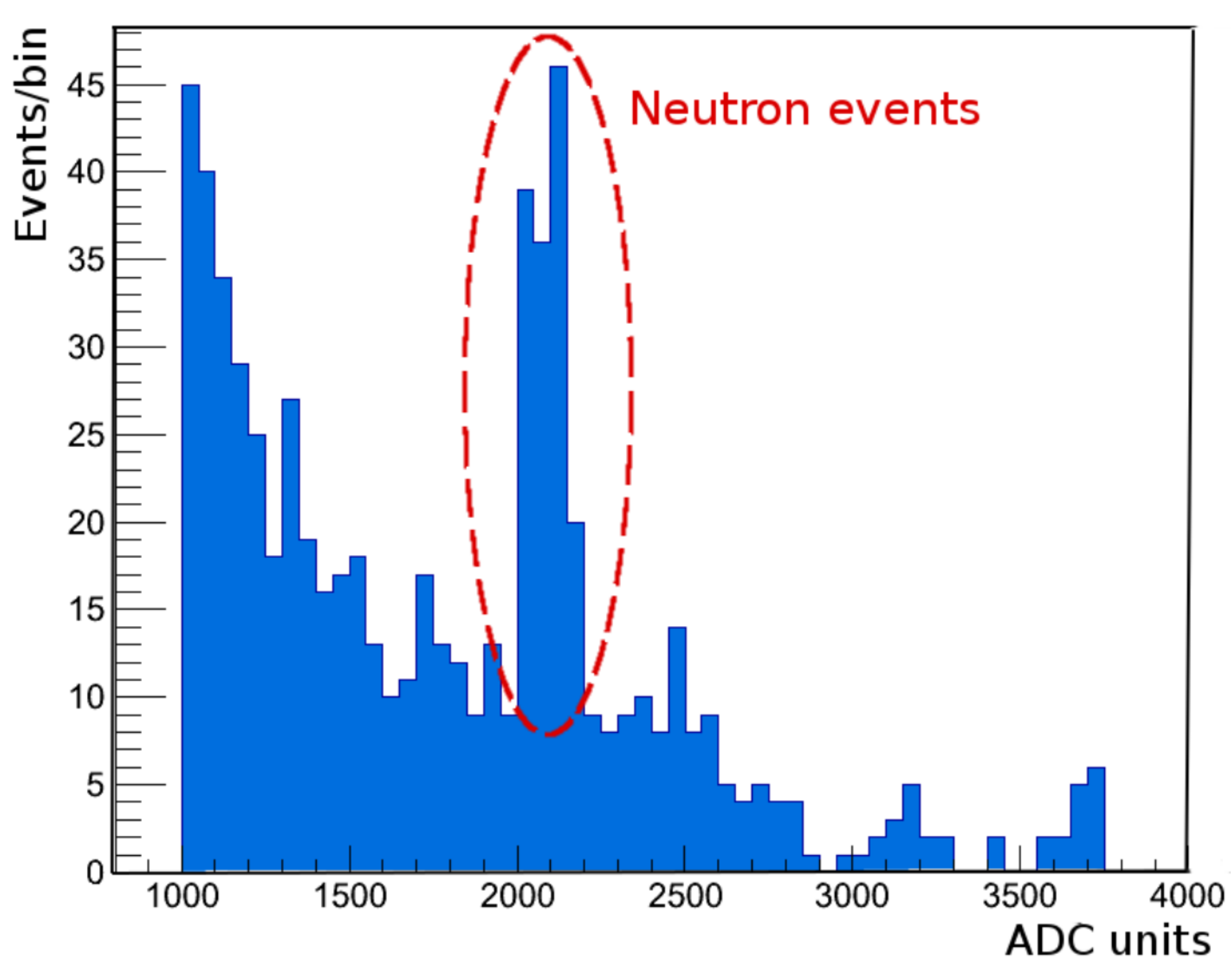}
  \end{center}
  \caption{The energy spectrum of room background with an exposure time of $\sim 1$ hour from a shielded neutron scintillator as used in the PoGOLino instrument. The signal peak from neutron capture events can clearly be distinguished from the exponential background spectrum.}
  \label{bg}
\end{figure}

Using both the sandwich configuration and the distinct spectral signature resulting from neutron capture by lithium, the background to the neutron detection becomes negligible. A typical room background spectrum acquired with an exposure time of $\sim 1$ hour using this setup can be seen in figure \ref{bg}. Despite the low signal rate of order $\sim 0.01\
\,\mathrm{Hz}$ the signal remains 
distinguishable from the mainly gamma induced background. These measurements show the capability of the relatively small neutron scintillators to perform statistically significant measurements of a low neutron flux. This property makes LiCAF scintillators suitable for usage on balloon flights during which the measurement environment changes rapidly. 

Measurements in two different energy bands are achieved by the use of neutron-moderating materials. The cross section for neutron capture by $^6\mathrm{Li}$ decreases rapidly with increasing neutron energy. By combining LiCAF with a moderator, the sensitive energy range can be broadened or shifted to higher energies, depending on the type and thickness of the moderator. This effect was simulated using Geant4 \cite{Geant4} for the combination of a polyethylene moderator and a LiCAF scintillator. Results are shown in figure \ref{energyresponse}. From this figure it can be seen that for an unshielded LiCAF scintillator (blue line), the detection efficiency is highest in the sub-eV energy range, however, when adding $70\,\mathrm{mm}$ of polyethylene to the detector (red line) this region is shifted to the $0.1-10\,\mathrm{MeV}$ region. For PoGOLino, it was decided to add a $70\,\mathrm{mm}$ moderator to one neutron scintillator while keeping the other unshielded and placing it at the maximum possible distance 
from the shielded detector. Thereby measuring in the thermal energy range with one detector while measuring over a broad range with the second.

The instrument contains a third scintillator, which is placed along with one of the neutron scintillators inside the polyethylene shield. This third scintillator has an identical geometry to the neutron detectors but instead of the LiCAF, a plastic scintillator of the same type as in PoGOLite, is used. This detector is used to study neutron scattering in plastic scintillators. Results from this scintillator will be discussed in a future paper.

The three PMTs are read out using a Flash Analog to Digital Converter (FADC) board which also performs the trigger and veto logic using programmable gate arrays. The FADC in turn is read out using a PC104 computer through a SpaceWire to Ethernet converter. Data downloading and communication with the PC104 proceeds through an ethernet-over-radio system (E-link) provided by the Esrange Space Center \cite{Esrange}. In case of a loss of communication, all data is also continuously stored on a solid state disk during flight. All detectors and electronics are placed in a pressure vessel containing a heating system in order to ensure both atmospheric pressure for the electronics and a relatively constant temperature for the detectors during ascent and float. Due to the temperature dependence on the thermalising properties of the moderating material and on the light yield of the BGO it is important to know the temperature in the pressure vessel during measurements. For this purpose three temperature sensors are placed inside the pressure vessel, these 
are read out using the PC104 computer. The total power consumption of the instrument is at maximum $20\,\mathrm{W}$ which is provided by a total of 8 batteries of type SAFT LSH20. The instrument was designed to start data taking automatically after connection of the batteries to the load using a external arming plug. The measurement time is limited by the batteries to approximately 6 hours, which exceeds the expected duration of prelaunch preparations and the flight itself. The total mass of the instrument is $13\,\mathrm{kg}$. A schematic of the full instrument design can be seen in figure \ref{instrument}.

The instrument was calibrated thoroughly before flight using gamma ($^{137}\mathrm{Cs}$ and $^{60}\mathrm{Co}$), beta ($^{90}\mathrm{Sr}$) and neutron ($^{252}\mathrm{Cf}$) sources in order to understand the scintillator response and the moderation effect of the polyethylene shield. Furthermore, long stability runs were performed, and the effect of temperature on the detector performance was studied in detail \cite{HT_2}.

\section{Flight}

\begin{figure}[h!]
  \centering
    \includegraphics[width=9.0 cm]{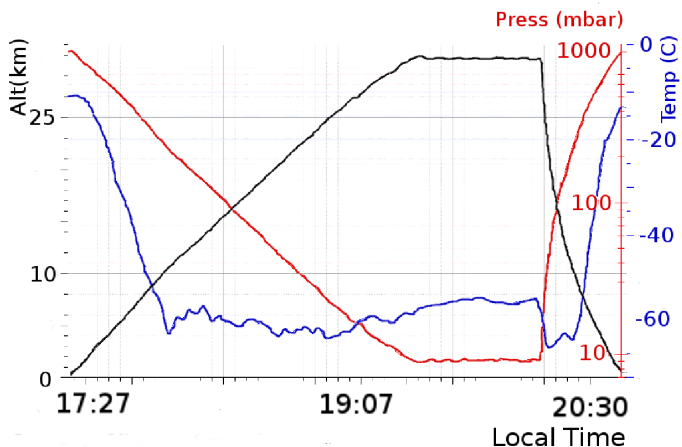}
  \caption{The altitude (black), pressure (red) and temperature (blue) as measured on the payload during the PoGOLino flight as a function of the local time. The altitude data was provided by $\cite{Kent}$.}
\label{flightdata}
\end{figure}

The launch of PoGOLino took place on March 20th 2013 from the Esrange Space Center in Northern Sweden at 17:20 local time. The altitude, along with the outside temperature and pressure during the flight, are shown in figure \ref{flightdata}. It can be seen that the instrument reached a maximum altitude of $30.9\,\mathrm{km}$ corresponding to an atmospheric overburden of $9\,\mathrm{g/cm^2}$. The outside temperature at this altitude was around $-60^\circ\,\mathrm{C}$. The temperature inside the pressure vessel throughout the flight was measured to be in the range of $0-5^\circ\mathrm{C}$.

\begin{figure}[h!]
  \centering
    \includegraphics[width=9.5 cm]{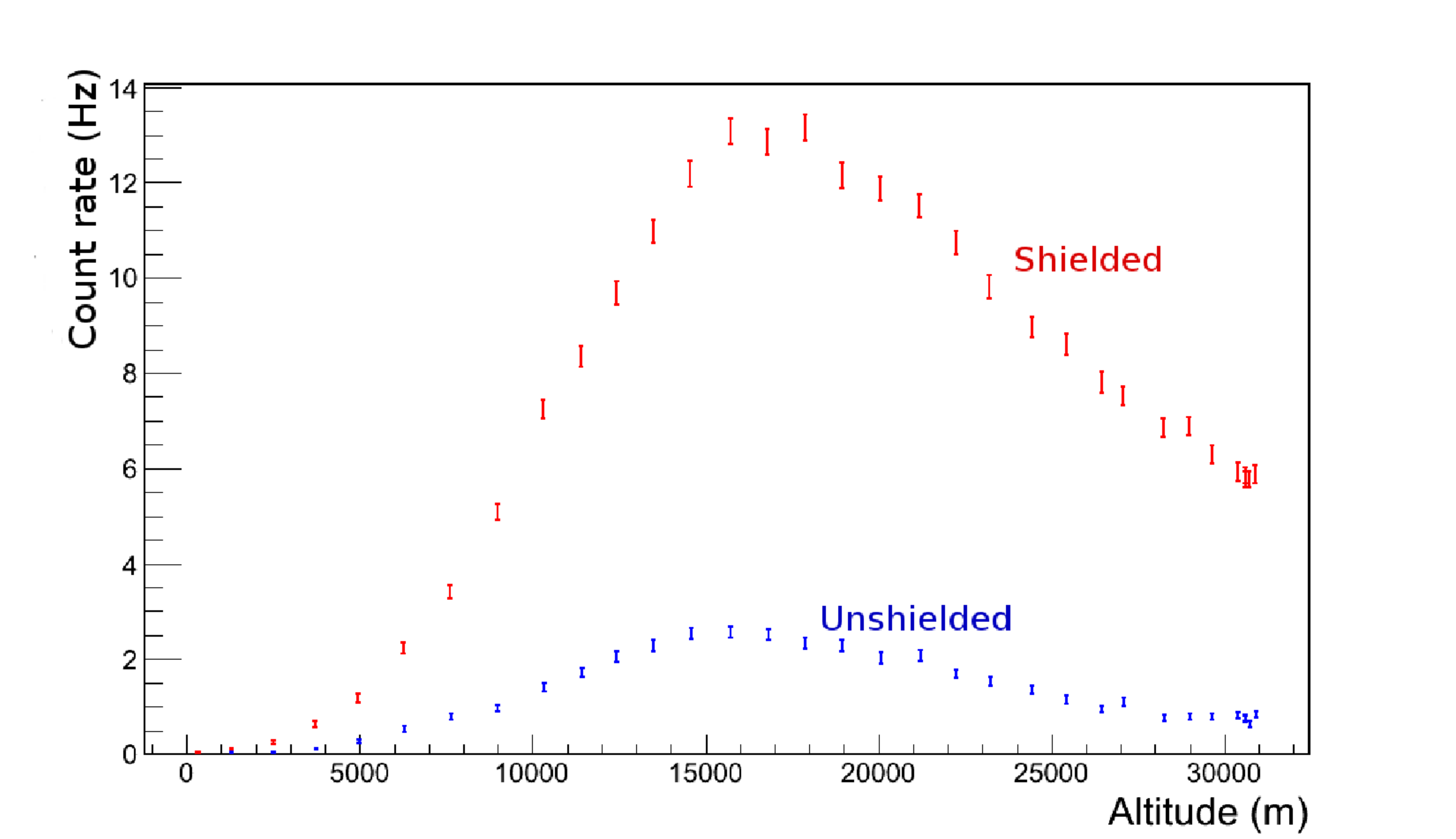}
  \caption{The count rate as measured in the shielded detector in red and that in the unshielded detector in blue plotted as a function of the altitude. Each data point is the result of a $300\,\mathrm{s}$ long measurement. The altitude data was provided by $\cite{Kent}$.}
\label{count_meas}
\end{figure}

The instrument performed as expected, and data were taken with all three detectors from one hour on ground prior to launch until 20:00 local time, when the payload was cut from the balloon and returned to ground by parachute. The count rates as measured by the two neutron scintillators can be seen in figure \ref{count_meas}. The figure shows that using the $5\,\mathrm{mm}$ thick scintillators sufficient statistics are acquired during the relatively short measurement sets of $300\,\mathrm{s}$ each. The results show the effect of the broadening of the sensitive range of the shielded detector by the polyethylene which results in higher rate in this detector with respect to the unshielded neutron scintillator. Furthermore one can see a maximum in the count rate corresponding to the altitude of the shower maximum.

\section{Simulations}

In order to understand the count rates as measured during the flight of PoGOLino and to be able to make estimates on the neutron induced background rates for the PoGOLite flight a series of simulations was performed. Simulations of the atmospheric neutron environment were performed using a version of PLANETOCOSMICS \cite{PLANETO} compatible with Geant4.9.5. This work resulted in simulated direction-dependent neutron energy spectra for different altitudes at all latitudes and longitudes and arbitrary solar activities. These spectra were then used as input for a series of Geant4 simulations of the PoGOLino gondola which resulted in the simulated counting rates for flight at three different altitudes. The PLANETOCOSMICS simulations will be discussed first followed by a description of the Geant4 simulations.

\subsection{PLANETOCOSMICS}

PLANETOCOSMICS is a Geant4 based simulation package which contains detailed models for the atmospheres and magnetic fields of different planets. For the work presented here the default model of the Earth (consisting fully of $\mathrm{SiO_2}$) was used together with the atmospheric model NRLMSISE00 \cite{NRM}, the height of the atmosphere was set to $100\,\mathrm{km}$. The internal magnetic field of the Earth was modelled using the IGRF model \cite{IGRF} whereas the external magnetic field and the magnetopause were modelled according to the TSY2001 model \cite{TSY}. The physics processes were handled using QGSP\_BIC\_HP \cite{QGSP} which uses the Binary model for energies between $20\,\mathrm{MeV}$ and $1\,\mathrm{GeV}$. Due to a shortcomings in the modelling capability of Geant4 with increasing Z of high energy nucleon interactions, all incoming cosmic rays were assumed to be protons for the work presented here. 

A set of simulation runs was performed each with a different proton energy range. The protons were emitted isotropically from a sphere with a radius of $2\times10^5\,\mathrm{km}$ centred around the Earth. The secondary particles produced in the atmosphere were sampled at all latitudes for altitudes of $11.3, 16, 30, 36, 40, 60$ and $99\,\mathrm{km}$. For all these secondary particles the energy and direction were stored to be used later to reproduce direction-dependent energy spectra at the different altitudes and latitudes. 

\begin{figure}[h!]
  \centering
    \includegraphics[width=9 cm]{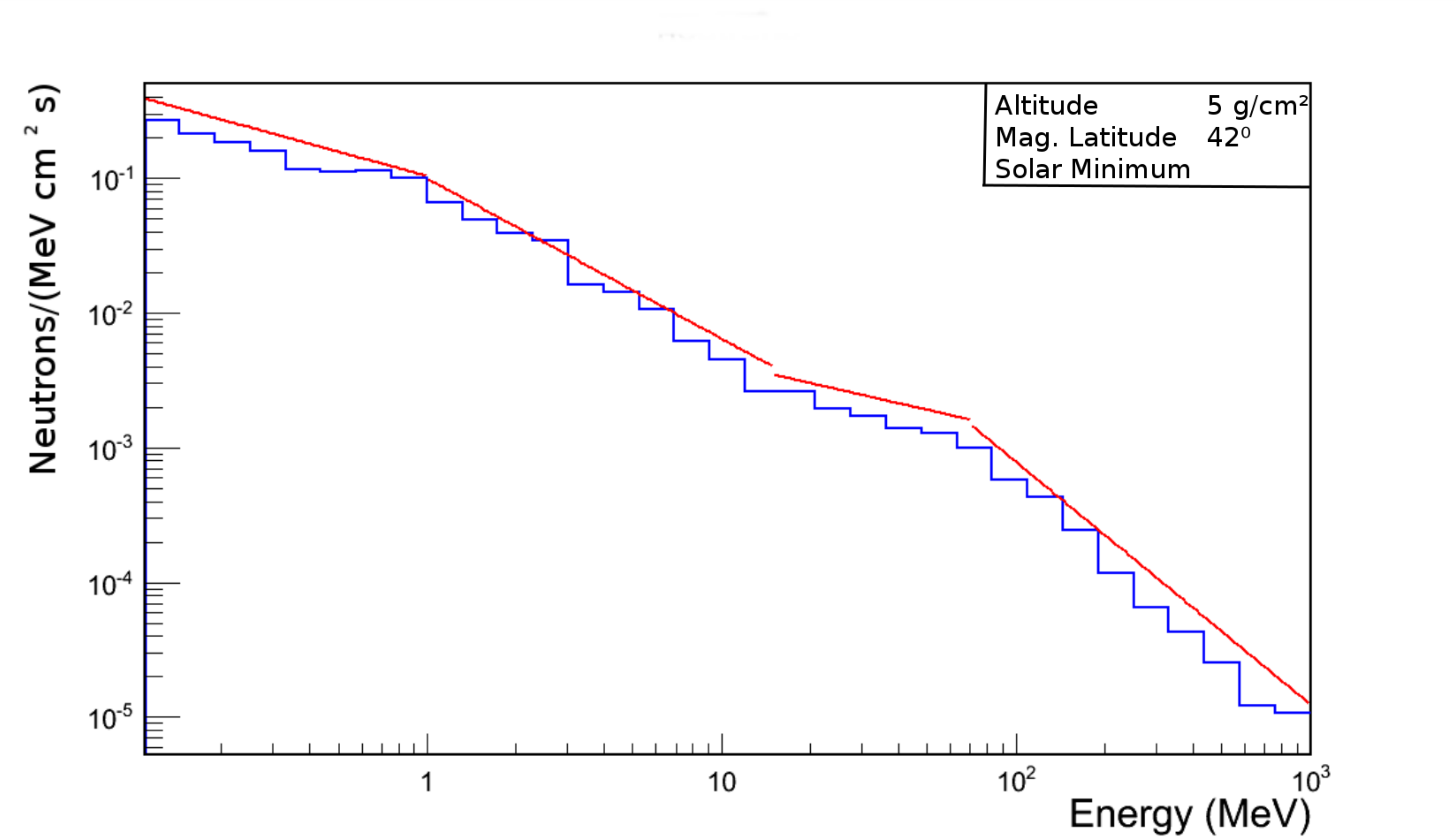}
  \caption{A comparison of the neutron spectrum as simulated by Armstrong et al. in \cite{Armstrong} (red line) and using the simulations described here (blue histogram).}
\label{Armstrong_vs_Kole}
\end{figure}

The set of simulations consisted of 11 runs for mono-energetic protons with energies of $1,2,3,4,5,6,7,8,9,10$ and $15\,\mathrm{GeV}$ and one run with a proton energy spectrum between $20$ and $\infty\,\mathrm{GeV}$ following a negative power law with an index of $2.71$ to resemble the cosmic ray energy spectrum. Different runs were used instead of one single measurement to give the possibility to recreate the final neutron spectra for different solar activities using one set of simulated data. The simulated proton and neutron spectra were renormalized using data of the proton spectra at the top of the atmosphere as measured by the PAMELA \cite{PAMELA} and BESS \cite{BESS} experiments for different solar activities. For example by normalizing on the basis of the PAMELA data of the proton spectrum during the last solar minimum \cite{PAMELA_meas}, the omnidirectional neutron spectrum at the latitude of Texas at an altitude equal to $5\,\mathrm{g/cm^2}$ was produced. The results are compared to that simulated in \cite{Armstrong} and are presented in figure \ref{Armstrong_vs_Kole}. From this figure it can be concluded that for the MeV energy range, the range in which most neutrons are produced, the results match fairly well. Differences partly stem from the use of different atmospheric models and incoming cosmic ray spectra.

\begin{figure}[h!]
  \centering
    \includegraphics[width=9 cm]{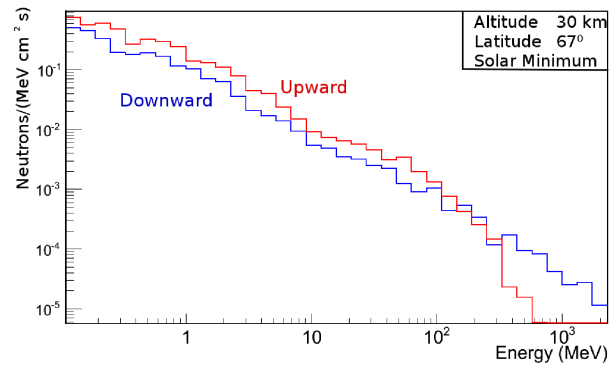}
  \caption{The neutron spectra as simulated using PLANETOCOSMICS for an altitude of $30\,\mathrm{km}$, latitude of $67^\circ$ during a solar minimum both for the upward moving neutrons in red and downward moving neutron in blue.}
\label{spectra}
\end{figure}

\begin{figure}[b!]
  \centering
    \includegraphics[width=9 cm]{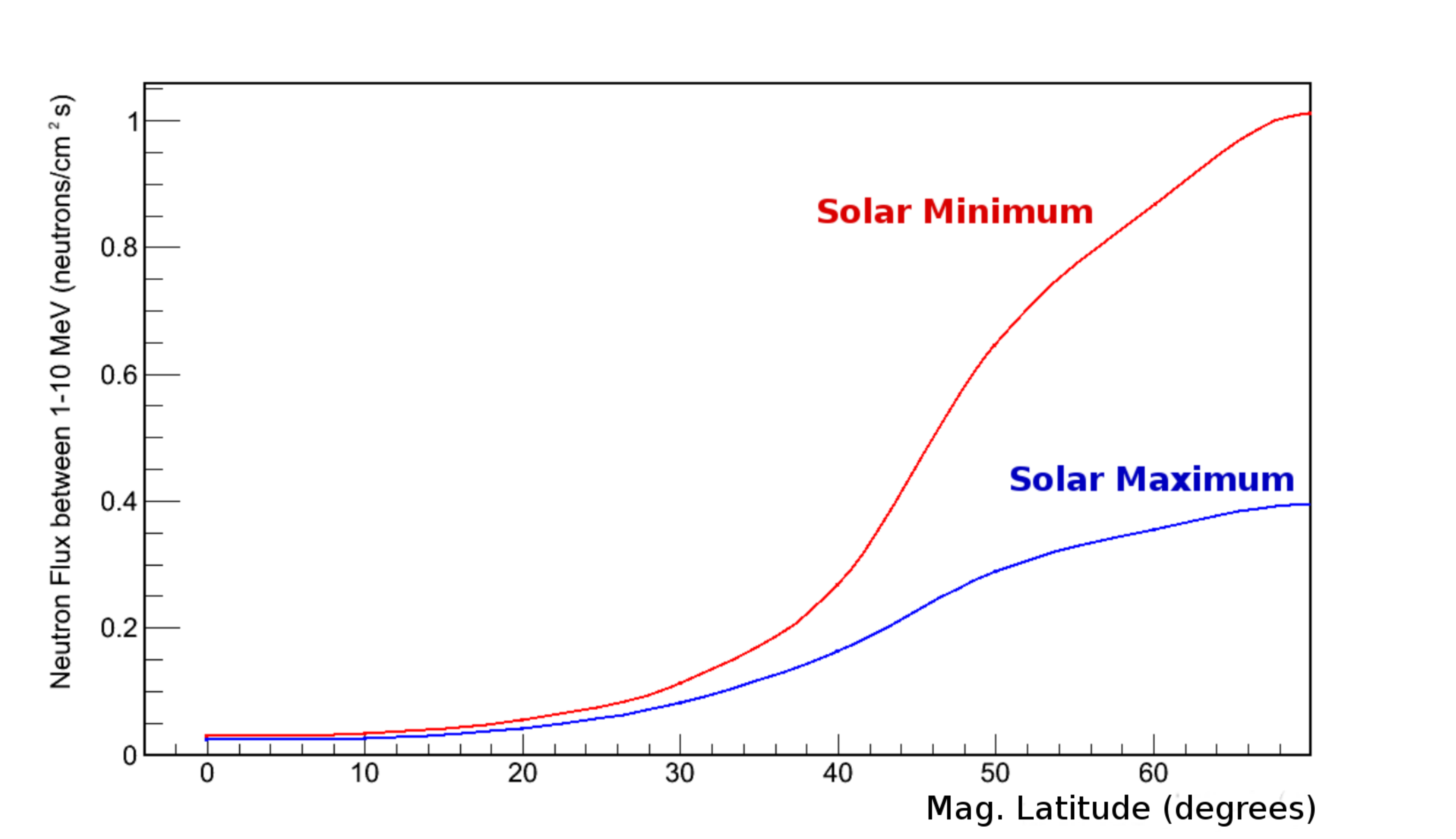}
  \caption{The integrated neutron flux between 1 and 10 MeV for an altitude of $30\,\mathrm{km}$ as a function of magnetic latitude both for a solar minimum and (in red) a solar maximum (in blue).}
\label{lat_effect}
\end{figure}

The same simulated data from PLANETOCOSMICS was used to produce the neutron spectra for solar maximum by normalizing the different runs using the proton spectra measured by the BESS experiment in 2002 \cite{BESS_maximum}. Finally, spectra were created for the solar activity which matches that of the PoGOLino flight. The neutron flux as measured by the Thule base of the Bartol Neutron Monitor \cite{Bartol} during the launch of PoGOLino matches that of the summer of 1999. For that summer data from the BESS experiment exists for the proton energy spectrum at the top of the atmosphere \cite{BESS_maximum}. This data was used to normalize the simulated data by PLANETOCOSMICS to create the neutron energy spectra at the Esrange Space Centre latitude for three different altitudes, $11.3, 16$ and $40\,\mathrm{km}$. An example of both the resulting upward and downward moving neutron spectra at an altitude of $30\,\mathrm{km}$ at the latitude of the Esrange Space Centre during a solar minimum can be 
seen in figure \ref{spectra}. It can be observed that due to the thermalisation in the atmosphere, the majority of which is below this altitude, the upward moving flux dominates for sub-GeV energies whereas for the unthermalised part above $1\,\mathrm{GeV}$ the downward moving flux dominates. To illustrate the effect of the latitude and solar activity on the atmospheric neutron environment the integrated neutron flux between 1 and 10 MeV as simulated using the described simulations method is shown as a function of the magnetic latitude in figure \ref{lat_effect}.

\subsection{Gondola Simulations}

\begin{figure}[t!]
  \centering
    \includegraphics[width=7 cm]{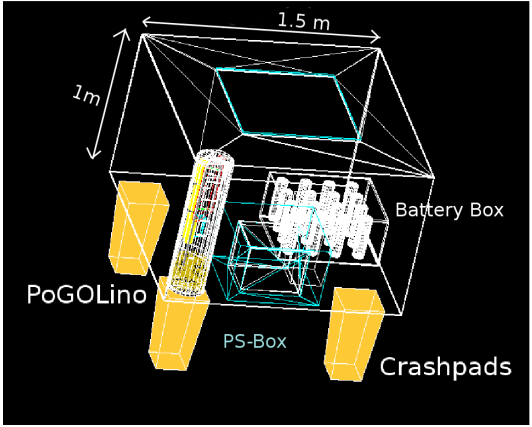}
  \caption{Example of the visualisation of the PoGOLino instrument as simulated in Geant4. The PoGOLino instrument is seen in the front left corner of the gondola. Next to it (in blue) one can see a polystyrene box containing unrelated electronics. Furthermore a large (32 kg) battery box, the cardboard crashpads and the gondola are taken into account in the simulations.}
\label{VI}
\end{figure}

The spectra simulated using PLANETOCOSMICS were used as input spectra for Geant4 simulations of a detailed model of the PoGOLino instrument and the gondola it was placed in. An example of the visualisation output can be seen in figure \ref{VI}. The detector setup contains the three PDCs, the polyethylene shield, the aluminium pressure vessel structure, a simplified representation of all the electronic components and the flight batteries. The Geant4 model of the instrument has been successfully verified using ground calibration data \cite{me}. The gondola geometry consists of the gondola structure, the crash-pads, a polystyrene box and its contents which was placed next to the detector and a large battery box which was present during the flight. In the simulation setup the neutrons were emitted from 2 separate hemispheres with a radius of $2\,\mathrm{m}$ centred around the gondola. The two separate hemispheres were used to account for the different spectra coming from above and below the instrument. The angular distribution with which the 
neutrons were emitted from the surfaces followed a cosine law, thus ensuring a uniform flux 
within 
the sphere. Valid events in the neutron scinitillators were selected by demanding the production of both a $\mathrm{^3T}$, with an energy in the range of $2.5-3\,\mathrm{MeV}$ and an $\alpha$ with an energy in the range of $1.8-2.2\,\mathrm{MeV}$. 

A similar set of Geant4 physics models was used for these simulations as for the PLANETOCOSMICS simulations. In order to correctly model sub-eV interactions the data library for thermal neutron scattering was also included. The temperature in the simulations was set to $278\,\mathrm{K}$, the temperature measured during the flight. 

The count rate for the two different LiCAF scintillators was simulated this way for three different altitudes. The results are shown in the following section where they are compared to the flight measurement results.

\section{Results}

\begin{figure}[h!]
  \centering
    \includegraphics[width=9 cm]{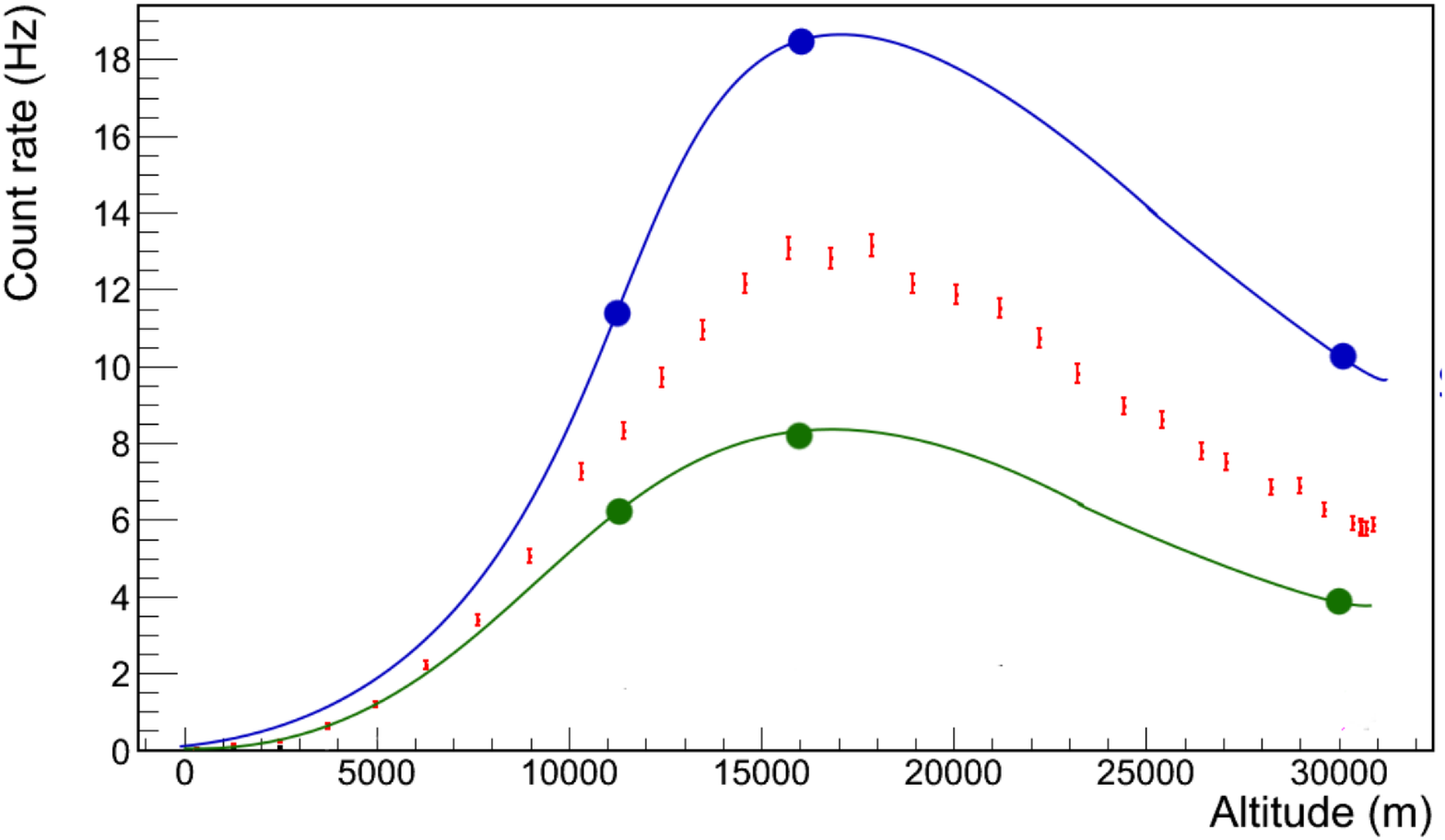}
  \caption{The count rate as measured by the shielded LiCAF detector in red as a function of the altitude compared to the simulated count rate for solar minimum (blue dots) and solar maximum (green dots). The blue and green line are placed to guide to eye.}
\label{maxmin_comp}
\end{figure}

\begin{figure}[h!]
  \centering
    \includegraphics[width=9 cm]{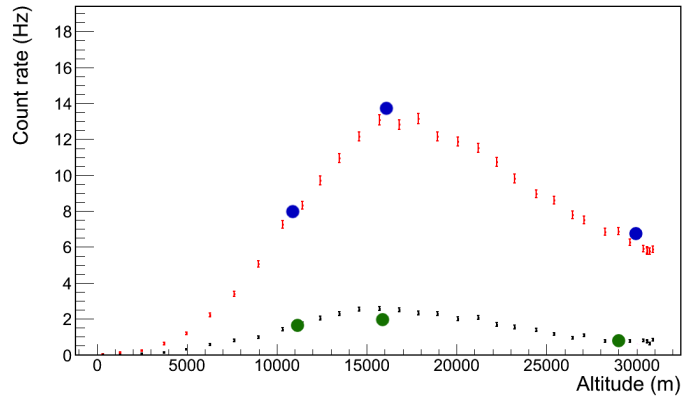}
  \caption{The count rate as measured by the shielded LiCAF detector in red and by the unshielded detector in black as a function of the altitude compared to the simulated rates (blue for the shielded and green for the unshielded detector).}
\label{Full_flight}
\end{figure}

In figure \ref{maxmin_comp} the measured count rate in the shielded detector is compared with the simulated rate for a solar minimum and a solar maximum flight. It can be seen that the measured rate is, as expected, between the two simulated extremes. In figure \ref{Full_flight} the measured data from both neutron scintillators is shown together with the simulated rates which make use of the proton spectra data as measured by the BESS experiment during the summer of 1999.  It can be concluded from this figure that the simulated count rate is in good agreement with the measured results. A small discrepancy is found for the unshielded detector, however, the simulated ratio  between the two scintillators is in good agreement with the measurements indicating that the spectral shape is correctly modelled. 

\section{Conclusion}

Measurements of neutrons at a high altitude and high latitude were successfully performed using BGO shielded LiCAF scintillators. These measurements show that LiCAF is a suitable material for performing high statistics neutron measurements within a relatively short period using a light weight detector. Furthermore we have shown that the measurements agree well with Geant4 based simulations, thereby validating the simulation procedure. We conclude that the results from the PLANETOCOSMICS simulations performed for the comparisons accurately describe the directionally dependent neutron energy spectra for high altitudes at different latitudes. The simulation method which makes use of PLANETOCOSMICS can be used to easily produce accurate neutron spectra for different locations at different times. We can use this method to simulate the neutron conditions for different instruments and applications at high altitudes. Specifically for the PoGOLite flight these spectra can be used to simulate the neutron induced background rates throughout the flight and thereby provide us with a better understanding of the polarization measurements as performed during the summer of 2013.


%

\section*{Acknowledgment}

The authors acknowledge: the Swedish National Space Board for funding the PoGOLino project. The Knut and Alice Wallenberg foundation for providing funds allowing Merlin Kole to attend the IEEE conference. The SSC Esrange Space Centre for their launch services and accepting PoGOLino on the TFB-01 test
flight. The AlbaNova workshop for their work on the PoGOLino mechanics. Laurent Desorgher for providing us with an unreleased version of PLANETOCOSMICS and finally Alex Howard and Christoffer Lundman for valuable discussion.

\ifCLASSOPTIONcaptionsoff
  \newpage
\fi

\end{document}